# Deep grey matter quantitative susceptibility mapping from small spatial coverages using deep learning

Xuanyu Zhu, Yang Gao, Feng Liu, Stuart Crozier, *Hongfu Sun

School of Information Technology and Electrical Engineering,
University of Queensland, Brisbane, Australia

***Correspondence**: Hongfu Sun
Address: Room 538, General Purpose South (Building 78),
University of Queensland, St Lucia QLD 4072, Australia
Email: hongfu.sun@uq.edu.au

**Abstract**

**Introduction:** Quantitative Susceptibility Mapping (QSM) is generally acquired with full brain coverage, even though many QSM brain-iron studies focus on the deep grey matter (DGM) region only. Reducing the spatial coverage to the DGM vicinity can substantially shorten the scan time or enhance the spatial resolution without increasing scan time; however, this may lead to significant DGM susceptibility underestimation.

**Method:** A recently proposed deep learning-based QSM method, namely xQSM, is investigated to assess the accuracy of dipole inversion on reduced brain coverages. The xQSM method is compared with two conventional dipole inversion methods using simulated and *in vivo* experiments from 4 healthy subjects at 3T. Pre-processed magnetic field maps are extended symmetrically from the centre of globus pallidus in the coronal plane to simulate QSM acquisitions of difference spatial coverages, ranging from 100% (~32 mm) to 400% (~128 mm) of the actual DGM physical size.

**Results:** The proposed xQSM network led to the lowest DGM contrast lost in both simulated and *in vivo* subjects, with the smallest susceptibility variation range across all spatial coverages. For the digital brain phantom simulation, xQSM improved the DGM susceptibility underestimation more than 20% in small spatial coverages, as compared to conventional methods. For the *in vivo* acquisition, less than 5% DGM susceptibility error was achieved in 48 mm axial slabs using the xQSM network, while a minimum of 112 mm coverage was required for conventional methods. It is also shown that the background field removal process performed worse in reduced brain coverages, which further deteriorated the subsequent dipole inversion.

**Conclusion**: The recently proposed deep learning-based xQSM method significantly improves the accuracy of DGM QSM from small spatial coverages as compared with conventional QSM algorithms, which can shorten DGM QSM acquisition time substantially.

# 1. Introduction

As an intrinsic tissue property, magnetic susceptibility can be used to detect and quantify disease biomarkers [1], such as iron, myelin, calcium, and hemorrhage [2]. In recent years an MRI post-processing technique, namely quantitative susceptibility mapping (QSM) [3], has been developed for measuring tissue magnetic susceptibility distribution *in vivo* [4]. QSM of the iron-rich deep grey matter (DGM) [5] has generated significant interest and has been investigated in multiple clinical studies, including Parkinson's disease [6-8], Huntington's disease [8], Alzheimer's disease [9, 10], and healthy aging [11]. However, QSM reconstruction from the raw MR phase data is non-trivial, which generally involves multi-channel receiver-coil combination [12], brain tissue extraction [13], phase unwrapping [14], background field removal [15-20] and dipole inversion [21-24].

Even though most brain iron studies focus on DGM only, one generally needs to acquire a substantially larger brain coverage for accurate QSM reconstruction in the DGM region. The reason is that the measured MR gradient-echo phase originates from the magnetic susceptibility source convoluting with a unit dipole kernel, which expands beyond its physical source location [25]. Thus, the relationship between the susceptibility source and the measured MR phase is non-local. A large field-of-view (FOV) coverage of the brain is required to perform the 3-dimensional dipole deconvolution robustly. A typical QSM gradient-echo sequence of 1 mm isotropic voxel size, with a parallel imaging factor of 2 applied for scan acceleration, takes 5 to 6 minutes to cover the whole brain [26]. One way to shorten the scan time is to use ultra-fast acquisition methods, such as 2D or 3D gradient-echo echo-planar imaging [27, 28], for full-brain QSM. There are also other acceleration techniques such as compressed sensing [29] and Wave-CAIPI [30], which could have better image quality with high acceleration factors. However, the image resolution and image qualify are generally compromised from these ultra-fast sequences. Alternatively, in theory, limiting the scan FOV to cover the DGM region only can reduce the scan time by 50-60% or enhance the spatial resolution without increasing scan time [31-33].

However, since the induced magnetic field extends beyond its physical susceptibility source substantially, reducing the FOV to the physical DGM region will inevitably truncate the non-local dipole field from the susceptibility source. Previous studies have reported significantly underestimated susceptibility measurements of DGM using conventional dipole inversion methods, such as MEDI [34] and TKD [35], on phase maps with small FOV coverages. Using numerical simulation and *in vivo* experiments, Elkady et al. [32] demonstrated that a minimum FOV covering 5.6 times the physical size of globus pallidus is required to maintain a <5% susceptibility error in globus pallidus, which is equivalent to a 76 mm axial slab. Another study from Karsa et al. [36] also found substantial susceptibility underestimation and contrast loss of DGM with FOV smaller than 50% of the full brain coverage due to insufficient dipole information for conventional QSM dipole inversion methods.

Deep learning, particularly convolutional neural networks (CNNs) [37], has received increasing attention as an alternative image reconstruction method in various medical imaging research areas. Deep learning-based methods capture the structure of the training dataset by adjusting learnable parameters, which improves the accuracy of the reconstruction and reduces the computation time. Recently, deep learning-based QSM methods have emerged. Bollman et al. [38] trained a full convolutional deep neural-network (DeepQSM) using synthetic datasets. Simple geometric structures were arranged randomly inside a cube to simulate the susceptibility sources as the training labels. The corresponding induced field maps were then calculated through the forward model as the training inputs. Another method, namely QSMnet, [39] proposed by Yoon et al., trained the neural network with *in vivo* dataset generated from multiple head orientations (i.e., COSMOS) [40]. Recently, Yang et al. [41] developed a new deep learning method, named xQSM, by introducing Octave convolution [42] into the U-net [43] architecture to improve the traditional convolutional layers. This Octave convolution network replaced the original convolution into four crossing operations, resulting in high and low-frequency groups of different matrix sizes, which saves storage and reduces the computation redundancy of traditional neural networks. It is shown that xQSM significantly improves DGM susceptibility estimation compared with other deep learning-based QSM methods [44]. This study thoroughly investigates the robustness of the deep learning-based xQSM dipole inversion on small FOV coverages for DGM susceptibility measurements, comparing its accuracy with conventional QSM dipole inversion methods.

## 2. Methods

### 2.1 Non-local magnetic dipole field

The magnetic field perturbation induced by a susceptibility source equals the convolution of the susceptibility distribution with the unit dipole kernel. Such a relationship can be simplified as multiplication in k-space:

$$F^{-1}DF\chi=\delta B$$
$$D = \frac{1}{3} - \frac{k_z^2}{k_x^2+k_y^2+k_z^2} \quad , \tag{1}$$

where $D$ represents the unit dipole kennel and $k_x$, $k_y$, $k_z$ represent the coordinates of the spatial frequency domain; $F$ and $F^{-1}$ denote the forward and inverse Fourier transforms; $\chi$ refers to the susceptibility map, and $\delta B = \Delta B/B_0$ is the field-strength normalized field map [45].

### 2.2 xQSM network structure and training datasets

Similar to the conventional U-net [43], the xQSM network [41] comprises two operations, i.e. encoding and decoding, referred to as down-sampling and up-sampling paths. The xQSM framework contains 10 Octave convolution layers (kernel size: 3×3×3), 2 max-pooling layers (kernel size: 2×2×2), 2 Octave transposed convolution layers (kernel size: 2×2×2), and 12 batch normalization layers, as shown in Figure 1(a). Rectified Linear Unit (ReLU) is adopted as the activation function of the network. The Octave convolution is explicitly designed to ensure that information in both resolution (i.e., high and low) feature groups communicate internally, which can be expressed as:

$$Y_H=\text{Conv}_{HH}(X_H)+\text{ConvT}(\text{Conv}_{LH}(X_L))$$

$$Y_L = \text{Conv}_{LL}(X_L) + \text{Conv}_{HL}(\text{AvgPool}(X_H)) \tag{2}$$

where Conv(·) represents traditional convolutions; ConvT(·) represents traditional transposed convolution operation of kernel size 2, doubling the dimensions of the low-resolution feature maps; AvgPool(·) is the average pooling operation, reducing the dimensions of the high-resolution feature maps to half. The Octave convolution operation is illustrated in Figure 1(b).

As detailed in the DeepQSM [38] and xQSM papers [41], training inputs (i.e., local field maps) were generated by convolving labels (i.e., susceptibility maps) with the unit dipole kernel according to the forward model in Eq. (1). Instead of using synthetic and simple geometries as training labels in DeepQSM, the xQSM adopts the more realistic human brain susceptibility maps LN-QSM [25] as the training labels for simulating the local field maps, which showed improved dipole inversion accuracy. This study extends the data generation in xQSM and compares two different training datasets, namely *Individual* and *Cropped*.

For the *Individual* mode, 96 *in vivo* LN-QSM volumes of size 144×192×128 were split into 15,000 individual QSM patches of size $64^3$ as the training labels. Local field maps, as the network inputs, were generated for the corresponding individual QSM patches by the dipole field forward calculation [45]. To avoid aliasing effect due to circular convolution, the forward calculation was slightly modified as: $\delta B = CF^{-1}DFP\chi$, where D represents the full-size dipole kernel; $\delta B$ and $\chi$ (both in size of $64^3$) are the simulated local field map and the corresponding QSM label used for training; *P* refers to a padding operation to increase the image size of the QSM label patch to full size by padding with zeros, while *C* is a joint cropping operation to extract the small-patch local field map from the simulated full-size field map.

For the *Cropped* mode, 96 full-size local field maps (i.e., 144×192×128) were calculated from the corresponding full-size susceptibility maps. A total of 15,000 field map patches (patch size $64^3$) were cropped from these 96 full-size local field maps, matching the susceptibility label patches in space as the network training inputs. Consequently, we named the above two trained networks as xQSM-Individual and xQSM-Cropped. For each training mode, the mean squared error (MSE) was adopted as the loss function. The training rate was set to $10^{-3}$. Each xQSM network training took about 18 hours (50 epochs) using 2 Tesla V100 GPUs with a minibatch size of 24.

### 2.3 Simulated and *in vivo* experiments

One healthy subject (27-year-old male) was scanned at 3T (Siemens Prisma) using a 3D unipolar 5-echo SWI (Susceptibility-Weighted Imaging [46]) sequence in 5 different head orientations, with parameters: 4.9 ms first TE, 4.9 ms echo spacing, 30 ms TR, 15° flip angle, 1 mm isotropic voxel, 144×192×128 mm$^3$ FOV, parallel imaging acceleration factor of 3, and total scan time of 4.3 min for each head orientation. The standard QSM pipeline was then processed, including brain mask generation using BET method [13], phase unwrapping using the best path method [14] and background field removal using the RESHARP method [16]. The COSMOS ("Calculation Of Susceptibility through Multiple Orientation Sampling") map from this subject was reconstructed as the ground truth label and used in the forward calculation model to simulate its local field map. To test the dipole inversion on high susceptibility sources, one hemorrhage (0.8 ppm of mean value and 0.1 ppm of standard deviation) and one calcification lesion (-0.4 ppm of mean value and 0.1 ppm of standard deviation) were simulated and added into this COSMOS data. This COSMOS subject with its simulated local field map is thus referred to as the simulation subject for method validation in the paper.

Another three healthy subjects (age: 28.3 ± 2.5 years) went through a 3D multi-echo gradient recalled-echo (GRE) scan using a 3T MR system (GE Discovery). Full brain GRE data were acquired in an axial slab with 8 unipolar readout echoes, 3.4 ms first TE, 3.5 ms echo spacing, 29.8 ms TR, 20° flip angle, 1 mm isotropic voxel, 256×256×128 mm$^3$ FOV, parallel acceleration factor of 2, and the total scan time of 5.9 min. One of these three subjects (26-year-old male) had full brain 3D GRE data acquired at five different head orientations to perform the COSMOS reconstruction as the QSM ground truth. This COSMOS subject was referred to as *in vivo* Subject #1, and the other two as *in vivo* Subject #2 (28-year-old female) and #3 (31-year-old male). For the three *in vivo* subjects, local field maps at neutral head positions were reconstructed through the established pre-processing pipeline, which is the same procedure used in the simulated subject.

**2.4 DGM coverages and susceptibility measurements**

To evaluate the effect of FOV coverages for DGM QSM, all local field maps were extended symmetrically from covering DGM only to the full brain in the coronal plane, as shown in Figure 2(a) and (b). Seven extended DGM coverages were tested, ranging from 100% (i.e., the minimum FOV) to 400% (i.e., maximum FOV, or full brain) of the actual DGM physical size. The minimum FOV, which is a 32 mm thick axial slab, covers the entire Globus Pallidus (GP), Putamen (PU), and Caudate Nucleus (CN). For the added hemorrhage and calcification susceptibility sources, seven coverages were extended symmetrically from 16 mm to 64 mm axial slabs. Dipole inversion was performed on the local field maps of different FOVs using xQSM, which was compared to another deep learning method DeepQSM [38] and two conventional methods iLSQR [47] and MEDI [34]. DGM regions-of-interest (ROIs) were drawn on each subject's central 5 axial QSM slices, containing the largest areas of DGM structures, as outlined in Figure 2(c) and (d). All DGM ROIs were drawn and measured using the ImageJ software, and the measurements were averaged from both brain hemispheres. All reported susceptibilities in this study were normalized relative to the chosen reference region Optic Radiation (OR) by subtraction.

Background field removal (BFR) is a critical QSM pre-processing step to generate local field maps without the overwhelming effect of the background field from air cavities. In this study, the RESHARP method [16] was used for BFR in the three *in vivo* experiments. Here we investigate the effects of BFR on dipole inversion of small FOV coverages by pre-processing the total field maps in different orders: (i) BFR then

Truncation (*B+T*, i.e., ideal case), (ii) Truncation then BFR (*T+B*, i.e., real-world acquisition), as outlined in Figure 3. The full-brain total field map processed firstly by BFR followed by FOV truncation (Figure 3(i)) could be regarded as the ideal case without the confounds of varied BFR performances on different FOVs. On the contrary, FOV truncation first on the total field map (Figure 3(ii)) represents the real-world acquisition scenario, which may affect the performance of the subsequent BFR process. The latter pipeline could lead to different local field maps and significantly affect the final QSM results. In this study, the purpose of investigating the ideal case is to mitigate the confounding effect from BFR so that one can isolate and examine the robustness of the dipole inversion process on its own.

## 3. Results
### 3.1 xQSM training modes comparison

QSM results of the minimum coverages (32 mm axial slab for DGM, 16 mm for hemorrhage and calcification) from the Simulation Subject are shown in Figure 4. DeepQSM, iLSQR and MEDI methods displayed significant susceptibility underestimation in DGM (yellow arrows in error maps), hemorrhage and calcification regions compared with the two xQSM methods. Moreover, the xQSM-Individual deep neural network resulted in less error than xQSM-Cropped, which is evident in the axial and coronal susceptibility error maps.

Mean susceptibility measurements of three DGM (i.e., GP, PU, and CN), one hemorrhage and one calcification regions from the Simulation Subject with different FOV coverages are plotted in Figure 5. Results from xQSM-Individual (red lines), xQSM-Cropped (blue lines) and DeepQSM (grey lines) are relatively stable along with different FOV coverages, while MEDI (green lines) and iLSQR measurements (yellow lines) spread a substantially larger susceptibility range. Among all methods, xQSM-Individual shows the best reconstruction results in the small FOV coverages of 100% and 150% DGM physical size, leading to the smallest errors (e.g., 3% and 6% underestimation for GP and PU in the minimum DGM coverage, respectively). The largest error is found in iLSQR, with over 40% susceptibility underestimation in the minimum DGM FOV. It can also be observed that xQSM of Individual training mode (i.e., xQSM-Individual, red lines) outperforms xQSM of Cropped training mode (i.e., xQSM-Cropped, blue lines) in all FOV coverages for the Simulation Subject. Thus the xQSM-Individual method is selected and referred to as the xQSM method for the *in vivo* experiments.

**3.2 Effect of background field removal for *in vivo* acquisitions**

QSM from both (i) BFR+Trucation ($B+T$) and (ii) Truncation+BFR ($T+B$) were quantified using *in vivo* Subject #1, with COSMOS serving as the ground truth. As compared in Figure 6, dipole inversion performs worse in $T+B$ (Figure 6(b)) than in $B+T$ (Figure 6(a)) for all methods. For example, the average DGM susceptibility underestimation of xQSM is only 4% in $B+T$ but 32% in $T+B$, for the minimum DGM FOV coverage. It is also observed that the deep-learning-based xQSM performs more robustly than the other two conventional dipole inversion methods in both cases. For the ideal case experiments (i.e., $B+T$, in 100% DGM coverage), the DeepQSM method shows the worst performance among all methods, with 31%, 42% and 36% susceptibility underestimation in GP, PU and CN, respectively. For real-world acquisition case (i.e., $T+B$), the MEDI method shows the worst performance among all methods on the same coverage, with 40%, 63% and 59% susceptibility underestimation in GP, PU and CN, respectively. Even though xQSM deviates substantially from the ground truth in the minimum DGM FOV in $T+B$ (Figure 6(b)), the errors decrease to under 5% from 150% DGM coverages onwards. Similar trends of improved susceptibility accuracy with larger FOVs are observed in conventional methods iLSQR and MEDI as well. However, unlike xQSM, substantial errors still present in 150% to 300% DGM coverages for MEDI and iLSQR methods, and the susceptibility values of DeepQSM were the lowest among all methods.

**3.3 Truncated QSM dipole inversion results from *in vivo* experiments**

Figure 7 displays the mean DGM QSM measurements and standard deviations for different FOV coverages from the three *in vivo* subjects. All experiments were performed with FOV truncation first, followed by RESHARP background field removal (i.e., *T+B*: real-world scenario). It is shown that xQSM led to the highest susceptibilities while DeepQSM the lowest across all FOV coverages. In general, the xQSM method maintains the smallest relative differences, and the MEDI method shows the largest relative differences in most FOV coverages, especially in smaller FOV coverages (i.e., 100% and 200% DGM physical size).

The relative QSM difference maps between smaller FOV coverages and the maximum full-brain coverage from *In vivo* Subject #1 are demonstrated in Figure 8. The full-brain results, normalized by their interquartile ranges, are shown in column (a), and relative difference maps from 100% DGM to the 350% DGM coverages are displayed from column (b) to (g). The four dipole inversion methods show no apparent difference in susceptibility contrast at the normalized full-brain coverage (column (a)), while the most pronounced underestimation occurs at 100% DGM coverage (column (b)), then gradually decreasing with larger spatial coverages. No apparent underestimation of DGM susceptibility appears in xQSM and DeepQSM results from coverages larger than 150% DGM physical size from visual inspection. Moreover, xQSM resulted in the least errors in 100% and 150% DGM sizes. In comparison, over 15% underestimation of DGM susceptibility remains in both MEDI and iLSQR results, particularly in 150% DGM FOVs (i.e., 48 mm axial slab). The reported NRMSEs of xQSM relative to the full-brain coverage exhibit the lowest deviation for all coverages among the three dipole inversion methods.

## 4. Discussion

In this study, we investigated the performance of susceptibility inversion on small FOV coverages, to substantially shorten the scan time, using a recently proposed learning-based xQSM method [41]. Four healthy subjects participated in the experiments, including one simulated local field map using the forward calculation and three post-processed local field maps from *in vivo* GRE phase measurements. We compared the mean susceptibilities of DGM regions from xQSM with another deep-learning method DeepQSM and two conventional dipole inversion methods (i.e., iLSQR and MEDI). It is observed that the deep learning-based xQSM method significantly improves the DGM QSM accuracy and stability in smaller FOV coverages for shortening scan time.

The previous DeepQSM [38], QSMnet [39], Auto-QSM [48], and QSMGAN [49] methods cropped the patches from full-size local field maps, which is the same as our xQSM-Cropped training mode. In this study, we also proposed the xQSM-individual training mode to crop the full-size susceptibility maps into small patches first, then forwardly simulate the local field maps of the individual small susceptibility patches. These two different training modes for xQSM were compared using our Simulation Subject experiment. The susceptibility results from xQSM *Individual* mode demonstrated the smallest error compared to xQSM Cropped mode and the conventional MEDI and iLSQR methods. The DGM susceptibility measurements from xQSM with *Individual* mode also showed the highest stability with the smallest measurement fluctuation with varying FOV coverages. The xQSM *Individual* mode performing better than *Cropped* mode may be because the individual susceptibility patches and their field maps from the forward calculation satisfy the full physics model of the dipole inversion. In contrast, in *Cropped* mode, this exact source-to-field relation is violated since the cropped field patch may contain dipole field from susceptibility sources outside of the patch region. The enhanced robustness of deep learning-based QSM methods against small FOVs originates from the convolution scheme of neural networks, also known as the local receptive field design. For each convolutional layer, a small localised local receptive field input was convoluted with the kernel, and thus the cropping operation only alters the edge but not the centre regions of the image, where DGM locate. By contrast, in conventional iterative methods, the truncation of the dipole field will lead to errors spreading the whole image space. Thus, a complete dipole field is essential for iterative dipole inversion methods.

We also investigated the effect of background field removal on the dipole field inversion by performing the background field removal process on the total field maps of different FOV coverages. Two MR phase pre-processing pipelines (i.e., different orders of performing background field removal and FOV truncation) were tested, one with background field removal on the total field maps first, followed by local field FOV truncations; the other one in reverse order. The former pre-processing order represents the ideal case that the same background field was removed for all FOV coverages and that the local field maps of different FOVs share the same overlapping region. In contrast, the latter represents the real-world scenario that background field removal was performed on total field maps of different FOV acquisitions, and the resulting local field maps may not be the same in their overlapping region. It is observed that the performance of the tested background field removal method (i.e., RESHARP) depends on the FOV coverages of the total field maps. The RESHARP background field removal combined with dipole inversion led to susceptibility underestimation of over 40% in GP and 50% in PU and CN using conventional MEDI and iLSQR methods in an acquisition coverage equal to DGM's physical size. The proposed xQSM reduces the susceptibility underestimation to 20% for FOV coverage of 100% DGM physical size and results in negligibly small deviations above 250% DGM coverages.

To summarize, we found that reducing spatial coverage will degrade all dipole inversion reconstruction methods since the dipole field from the DGM source extends beyond its physical space and has a critical impact on QSM reconstruction accuracy. From the analysis of simulation and *in vivo* experiments, it is concluded that an acquisition slab covering only the DGM region (~32 mm) is inadequate for accurate xQSM reconstruction, resulting in at least a 20% underestimation of mean susceptibility. The minimum coverage required for the learning-based xQSM method needs to be at least 150% (i.e. 48 mm) of the physical dimension of the DGM region to keep the error under 5%. While maintaining an error within 5%, traditional iterative methods MEDI and iLSQR need 350% (i.e., 112 mm) of the extended physical dimension of DGM. It means that by using xQSM for dipole inversion of the DGM, we can substantially reduce the scan time. The better performance of xQSM under reduced FOV coverages compared to conventional dipole fitting methods (i.e., MEDI and iLSQR) maybe because the trained xQSM neural network has learned the pattern between susceptibility source and truncated dipole field from the small patches. In this study, three different training patch sizes (i.e., $48^3$, $64^3$, and $80^3$) were compared to investigate the effects of the training patch size on the network's performance under small FOV coverages. As shown in Supp. Fig. 1 and Supp. Table 1, all three networks achieved comparable reconstructions under small FOV coverages, with the network trained with the patch size of $64^3$ leading to the best results.

There are a few limitations in this study. First, our experiments were performed only on the image resolution of 1 mm isotropic, which is the same as for xQSM data training. The reconstruction accuracy of xQSM for different resolutions needs to be further investigated. Second, only three regions of interest were measured, including globus pallidus, putamen and caudate nucleus. Other iron-rich regional substructures, such as the thalamus, substantia nigra, and red nucleus, will require larger coverages than the current 48 mm axial slab. Third, both deep learning methods in this paper, i.e., DeepQSM [38] and xQSM [41], are limited to phase data that are acquired in the axial orientation, similar as QSMnet [39]. Pre and post image rotations can be used for other head orientation configurations. However, this will introduce image blurring due to interpolation, which requires further development of deep learning QSM in the future. Finally, this study shows deteriorated background field removal (i.e., RESHARP) performance at smaller spatial coverages, which further confounded the truncated dipole inversion. Future studies will investigate different background field removal methods to mitigate this effect. Alternatively, end-to-end deep learning-based methods could be developed for total field inversion, such as autoQSM [48], to eliminate the error propagation through the background field removal process.

**5. Conclusion**

A recently proposed deep learning-based QSM method (xQSM) is investigated for DGM QSM on localized FOV coverages, which suggests great potential to shorten the QSM scan time substantially. The xQSM method was tested in low spatial coverages on both simulated and *in vivo* datasets and compared with a deep learning (i.e. DeepQSM) and two conventional algorithms (i.e., iLSQR and MEDI). Only less than 5% DGM susceptibility error was achieved in 48 mm axial slab coverage using the xQSM method, while a minimum of 112 mm coverage was required for other methods.

**Reference**

1. Duyn, J.H. and J. Schenck, *Contributions to magnetic susceptibility of brain tissue.* NMR in Biomedicine, 2017. **30**(4): p. e3546.
2. Wang, Y. and T. Liu, *Quantitative susceptibility mapping (QSM): decoding MRI data for a tissue magnetic biomarker.* Magnetic resonance in medicine, 2015. **73**(1): p. 82-101.
3. Haacke, E.M., et al., *Quantitative susceptibility mapping: current status and future directions.* Magnetic resonance imaging, 2015. **33**(1): p. 1-25.
4. Sun, H., et al., *Validation of quantitative susceptibility mapping with Perls' iron staining for subcortical gray matter.* Neuroimage, 2015. **105**: p. 486-492.
5. Lim, I.A.L., et al., *Human brain atlas for automated region of interest selection in quantitative susceptibility mapping: application to determine iron content in deep gray matter structures.* Neuroimage, 2013. **82**: p. 449-469.
6. Dexter, D., et al., *Alterations in the levels of iron, ferritin and other trace metals in Parkinson's disease and other neurodegenerative diseases affecting the basal ganglia.* Brain, 1991. **114**(4): p. 1953-1975.
7. Acosta-Cabronero, J., et al., *The whole-brain pattern of magnetic susceptibility perturbations in Parkinson's disease.* Brain, 2017. **140**(1): p. 118-131.
8. Chen, J.C., et al., *MR of human postmortem brain tissue: correlative study between T2 and assays of iron and ferritin in Parkinson and Huntington disease.* American journal of neuroradiology, 1993. **14**(2): p. 275-281.
9. LeVine, S.M., *Iron deposits in multiple sclerosis and Alzheimer's disease brains.* Brain research, 1997. **760**(1-2): p. 298-303.
10. Bouras, C., et al., *A laser microprobe mass analysis of brain aluminum and iron in dementia pugilistica: comparison with Alzheimer's disease.* European neurology, 1997. **38**(1): p. 53-58.
11. Betts, M.J., et al., *High-resolution characterisation of the aging brain using simultaneous quantitative susceptibility mapping (QSM) and R2\* measurements at 7 T.* Neuroimage, 2016. **138**: p. 43-63.

12. Schweser, F., et al., *Quantitative imaging of intrinsic magnetic tissue properties using MRI signal phase: an approach to in vivo brain iron metabolism?* Neuroimage, 2011. **54**(4): p. 2789-2807.
13. Cronin, M.J., et al., *Exploring the origins of echo-time-dependent quantitative susceptibility mapping (QSM) measurements in healthy tissue and cerebral microbleeds.* Neuroimage, 2017. **149**: p. 98-113.
14. Abdul-Rahman, H.S., et al., *Fast and robust three-dimensional best path phase unwrapping algorithm.* Applied optics, 2007. **46**(26): p. 6623-6635.
15. Schweser, F., et al., *An illustrated comparison of processing methods for phase MRI and QSM: removal of background field contributions from sources outside the region of interest.* NMR in Biomedicine, 2017. **30**(4): p. e3604.
16. Sun, H. and A.H. Wilman, *Background field removal using spherical mean value filtering and Tikhonov regularization.* Magnetic resonance in medicine, 2014. **71**(3): p. 1151-1157.
17. Wang, S., et al., *Noise effects in various quantitative susceptibility mapping methods.* IEEE Transactions on Biomedical Engineering, 2013. **60**(12): p. 3441-3448.
18. Liu, T., et al., *A novel background field removal method for MRI using projection onto dipole fields.* NMR in Biomedicine, 2011. **24**(9): p. 1129-1136.
19. Zhou, D., et al., *Background field removal by solving the Laplacian boundary value problem.* NMR in Biomedicine, 2014. **27**(3): p. 312-319.
20. Groetsch, C., *The theory of tikhonov regularization for fredholm equations.* 104p, Boston Pitman Publication, 1984.
21. Polak, D., et al., *Nonlinear dipole inversion (NDI) enables robust quantitative susceptibility mapping (QSM).* NMR Biomed, 2020. **33**(12): p. e4271.
22. Robinson, S.D., et al., *An illustrated comparison of processing methods for MR phase imaging and QSM: combining array coil signals and phase unwrapping.* NMR in Biomedicine, 2017. **30**(4): p. e3601.
23. Milovic, C., et al., *Fast nonlinear susceptibility inversion with variational regularization.* Magnetic resonance in medicine, 2018. **80**(2): p. 814-821.
24. Sun, H., et al., *Quantitative susceptibility mapping using a superposed dipole inversion method: application to intracranial hemorrhage.* Magnetic resonance in medicine, 2016. **76**(3): p. 781-791.
25. Sun, H., et al., *Whole head quantitative susceptibility mapping using a least-norm direct dipole inversion method.* NeuroImage, 2018. **179**: p. 166-175.
26. Kruse, S., et al. *Fast EPI based 3D MR elastography of the brain.* in *Proceedings of the International Society for Magnetic Resonance in Medicine.* 2006.
27. Langkammer, C., et al., *Fast quantitative susceptibility mapping using 3D EPI and total generalized variation.* Neuroimage, 2015. **111**: p. 622-630.
28. Le Ster, C., et al., *Comparison of SMS-EPI and 3D-EPI at 7T in an fMRI localizer study with matched spatiotemporal resolution and homogenized excitation profiles.* Plos one, 2019. **14**(11): p. e0225286.
29. Lustig, M., D. Donoho, and J.M. Pauly, *Sparse MRI: The application of compressed sensing for rapid MR imaging.* Magn Reson Med, 2007. **58**(6): p. 1182-95.

30. Bilgic, B., et al., *Wave-CAIPI for highly accelerated 3D imaging.* Magn Reson Med, 2015. **73**(6): p. 2152-62.
31. Bilgic, B., et al. *Single-step QSM with fast reconstruction.* in *Third International Workshop on MRI Phase Contrast & Quantitative Susceptibility Mapping.* 2014.
32. Elkady, A.M., H. Sun, and A.H. Wilman, *Importance of extended spatial coverage for quantitative susceptibility mapping of iron-rich deep gray matter.* Magnetic resonance imaging, 2016. **34**(4): p. 574-578.
33. Liu, Z., et al., *Preconditioned QSM to determine a large range of susceptibility over the entire field of view from total field.* Proc Int SocMagn Reson Med, 2016. **24**: p. 0032.
34. Liu, J., et al., *Morphology enabled dipole inversion for quantitative susceptibility mapping using structural consistency between the magnitude image and the susceptibility map.* Neuroimage, 2012. **59**(3): p. 2560-2568.
35. Shmueli, K., et al., *Magnetic susceptibility mapping of brain tissue in vivo using MRI phase data.* Magnetic Resonance in Medicine: An Official Journal of the International Society for Magnetic Resonance in Medicine, 2009. **62**(6): p. 1510-1522.
36. Karsa, A., S. Punwani, and K. Shmueli, *The effect of low resolution and coverage on the accuracy of susceptibility mapping.* Magnetic resonance in medicine, 2019. **81**(3): p. 1833-1848.
37. Xu, L., et al., *Deep convolutional neural network for image deconvolution.* Advances in neural information processing systems, 2014. **27**: p. 1790-1798.
38. Bollmann, S., et al., *DeepQSM - using deep learning to solve the dipole inversion for quantitative susceptibility mapping.* Neuroimage, 2019. **195**: p. 373-383.
39. Yoon, J., et al., *Quantitative susceptibility mapping using deep neural network: QSMnet.* Neuroimage, 2018. **179**: p. 199-206.
40. Liu, T., et al., *Calculation of susceptibility through multiple orientation sampling (COSMOS): a method for conditioning the inverse problem from measured magnetic field map to susceptibility source image in MRI.* Magnetic Resonance in Medicine: An Official Journal of the International Society for Magnetic Resonance in Medicine, 2009. **61**(1): p. 196-204.
41. Gao, Y., et al., *xQSM: quantitative susceptibility mapping with octave convolutional and noise-regularized neural networks.* NMR in Biomedicine, 2021. **34**(3): p. e4461.
42. Chen, Y., et al. *Drop an octave: Reducing spatial redundancy in convolutional neural networks with octave convolution.* in *Proceedings of the IEEE/CVF International Conference on Computer Vision.* 2019.
43. Ronneberger, O., P. Fischer, and T. Brox. *U-net: Convolutional networks for biomedical image segmentation.* in *International Conference on Medical image computing and computer-assisted intervention.* 2015. Springer.
44. Schweser, F., et al., *Quantitative susceptibility mapping for investigating subtle susceptibility variations in the human brain.* Neuroimage, 2012. **62**(3): p. 2083-2100.
45. Liu, C., et al., *Quantitative susceptibility mapping: contrast mechanisms and clinical applications.* Tomography, 2015. **1**(1): p. 3.

46. Haacke, E.M., et al., *Susceptibility weighted imaging (SWI).* Magnetic Resonance in Medicine: An Official Journal of the International Society for Magnetic Resonance in Medicine, 2004. **52**(3): p. 612-618.
47. Li, W., et al., *A method for estimating and removing streaking artifacts in quantitative susceptibility mapping.* Neuroimage, 2015. **108**: p. 111-122.
48. Wei, H., et al., *Learning-based single-step quantitative susceptibility mapping reconstruction without brain extraction.* NeuroImage, 2019. **202**: p. 116064.
49. Chen, Y., et al., *QSMGAN: improved quantitative susceptibility mapping using 3D generative adversarial networks with increased receptive field.* NeuroImage, 2020. **207**: p. 116389.

**Figures and Captions**

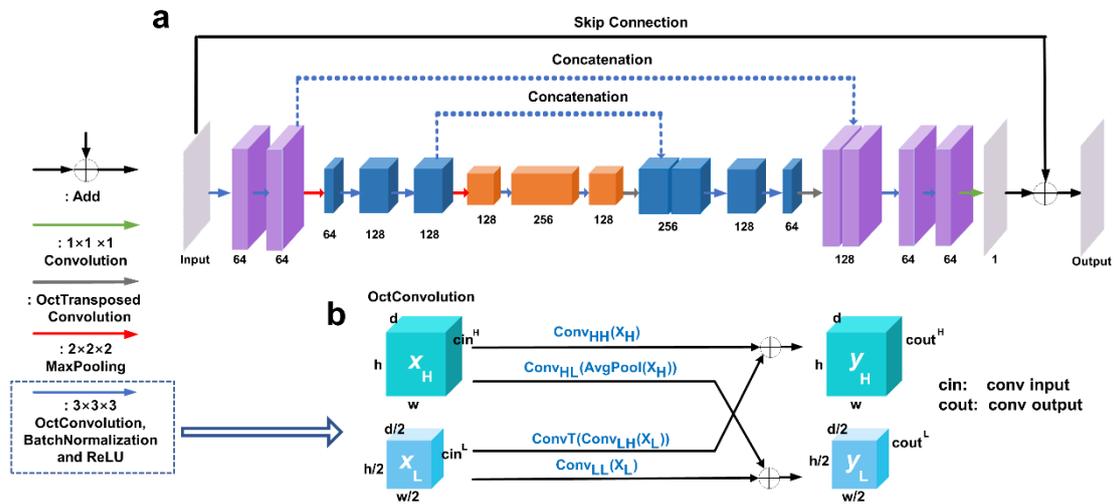

**Figure 1.** (a) The xQSM neural network structure, incorporating the new Octave convolution on the U-net architecture. (b) The cross-convoluted scheme of the new OctConvolution operation, with each selected feature divided into high (H) and low (L) groups.

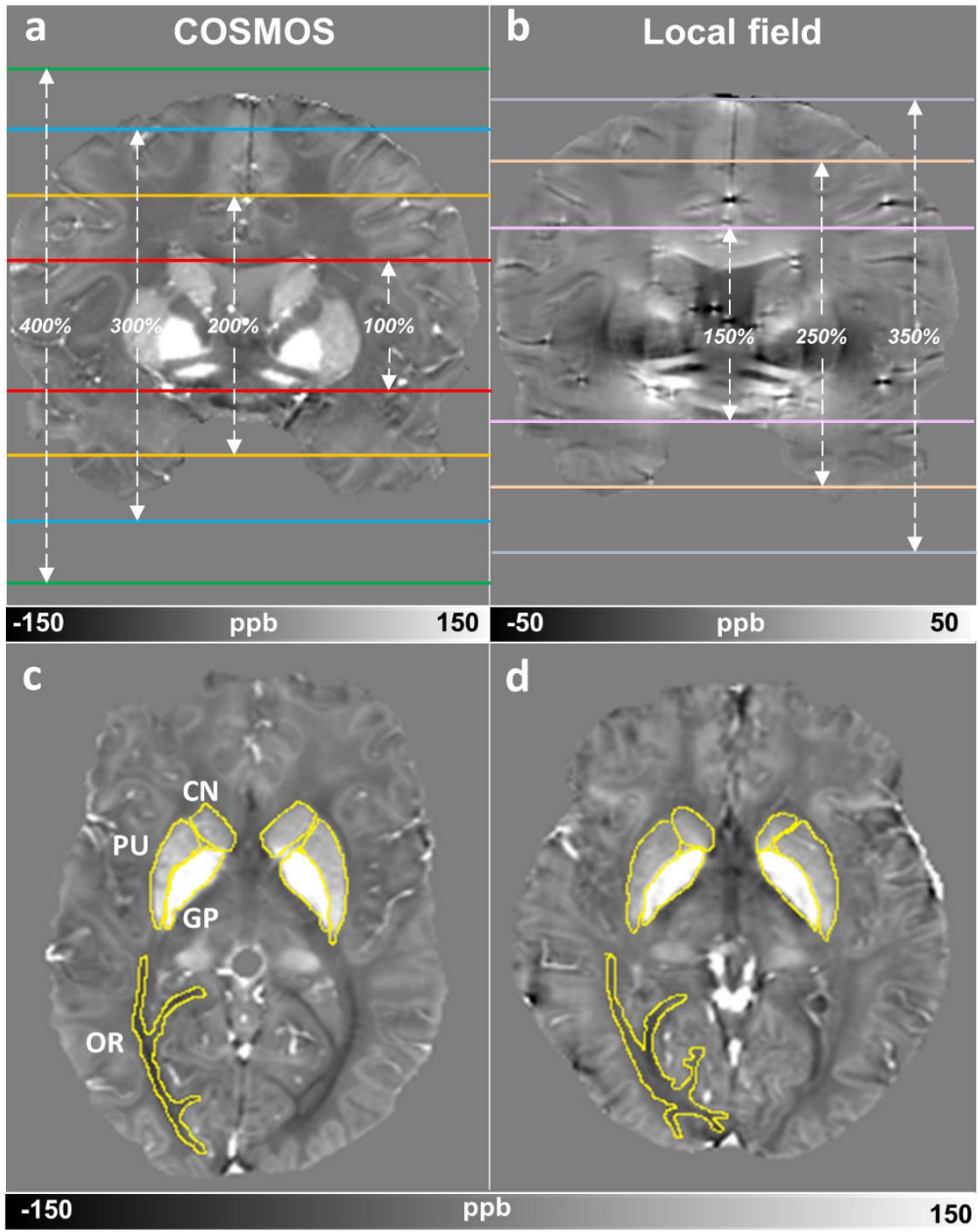

**Figure 2**. Different extended DGM coverages (blue outlined in different color lines) illustrated on QSM (a) and local field map (b) in coronal view. DGM ROIs (outlined in yellow), including globus pallidus (GP), caudate nucleus (CN) and putamen (PU) were drawn on the Simulation Subject (c) and in vivo Subject #1 (d).

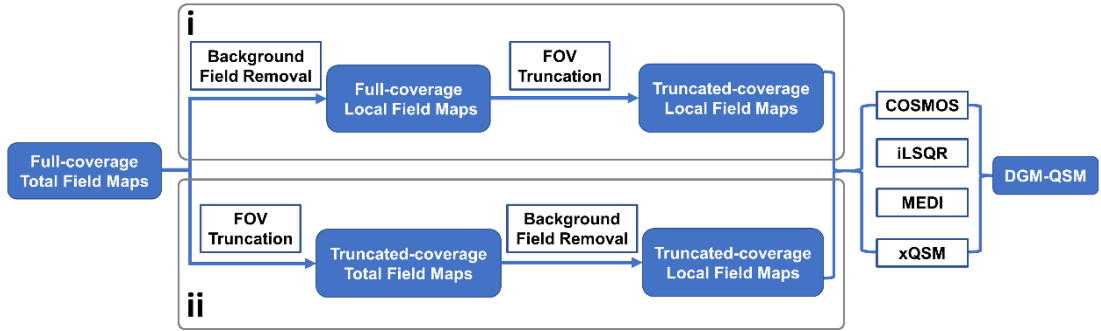

**Figure 3.** Two QSM pre-processing pipelines of BFR and FOV truncations. (i) Ideal case with the same BFR performed on different FOV truncations. (ii) Actual real-world case with BFR performed on the total field maps of different FOV coverages.

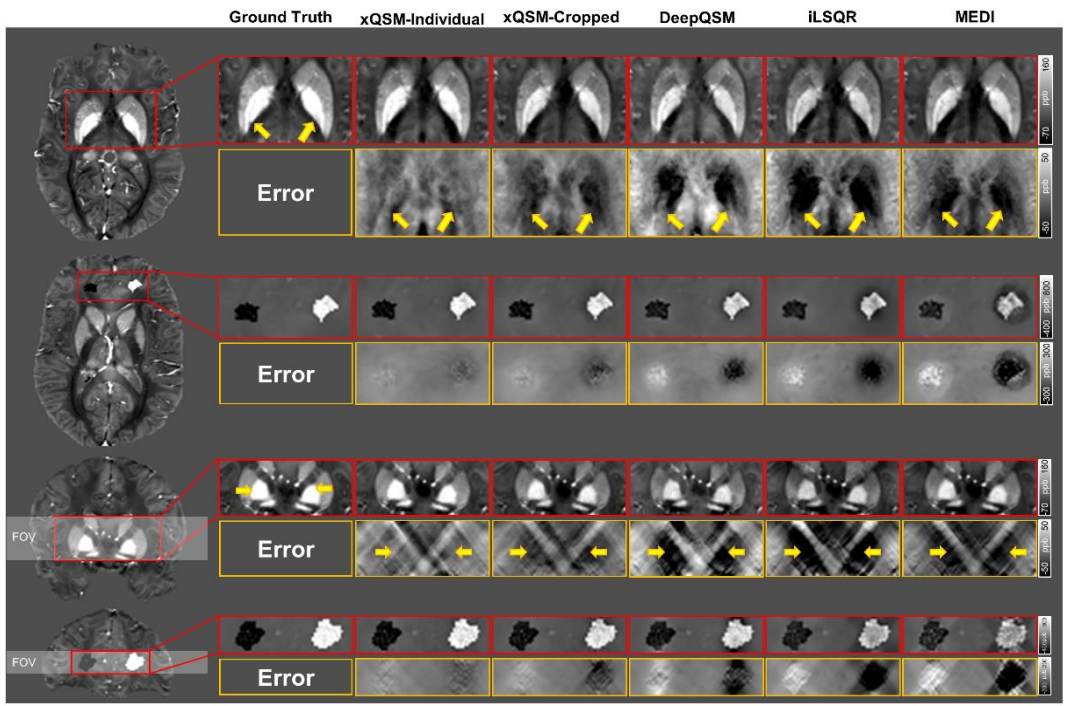

**Figure 4.** QSM results of DGM (globus pallidus, putamen and caudate nucleus), hemorrhage and calcification from four dipole inversion methods, including two xQSM (i.e., Individual and Cropped training modes), DeepQSM and two conventional (iLSQR and MEDI) methods, on the minimum FOV coverages. Error maps relative to the ground truth are shown in the bottom rows, and substantial susceptibility underestimations in ROIs are pointed by yellow arrows. The grey shaded area on coronal susceptibility map indicates the minimum FOV coverage.

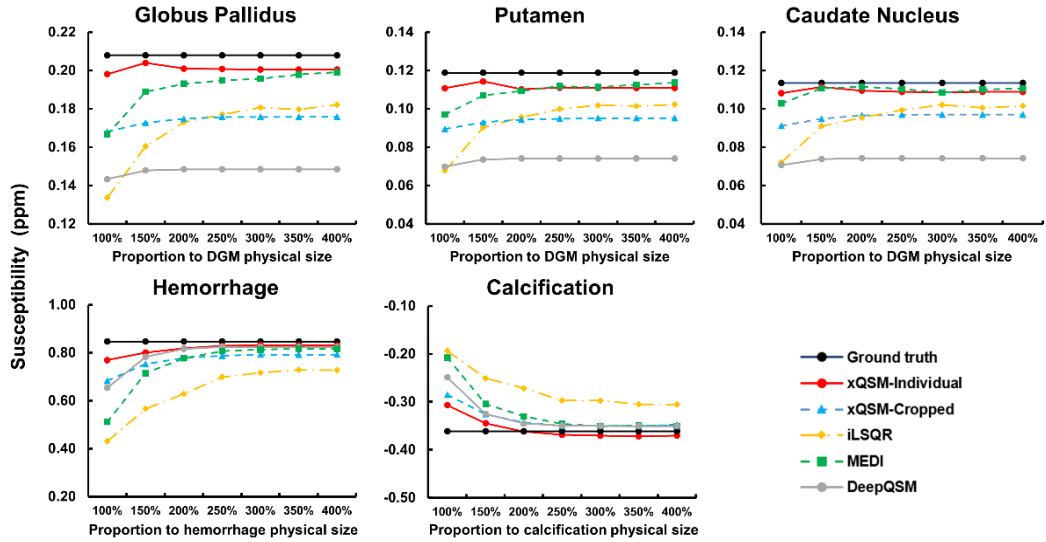

**Figure 5.** DGM (globus pallidus, putamen and caudate nucleus), hemorrhage and calcification measurements of the Simulation Subject from different FOV coverages compared among the five different QSM dipole inversion methods.

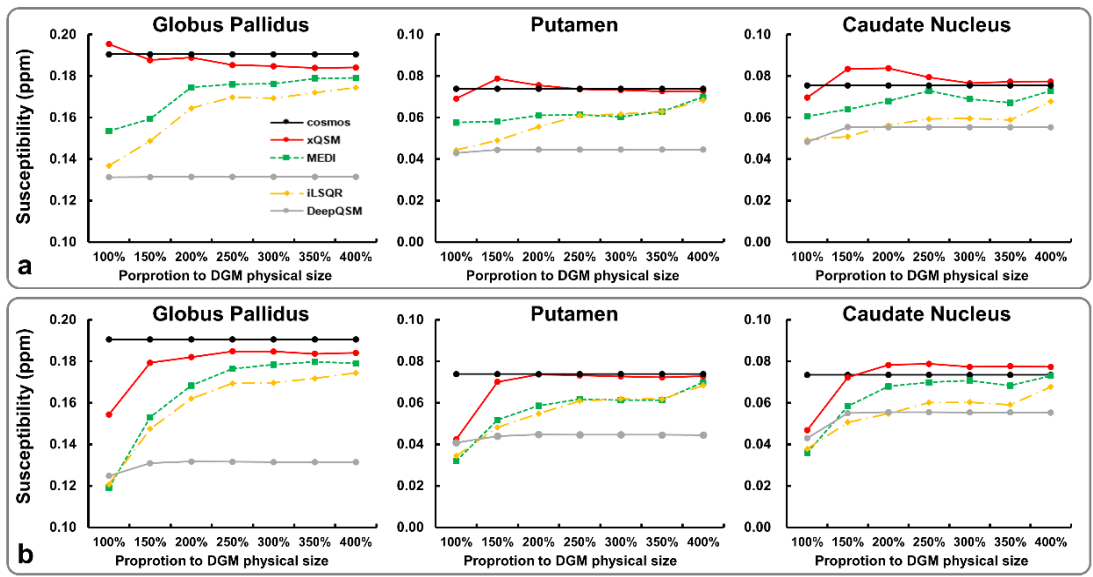

**Figure 6.** DGM (globus pallidus, caudate nucleus and putamen) measurements from in vivo Subject #1 with different FOVs compared between xQSM, DeepQSM, MEDI, iLSQR, and COSMOS. Different orders of background field removal and FOV truncation are compared in (a) BFR then Truncation (B+T, i.e., ideal case) and (b) Truncation then BFR (T+B, i.e., real-world acquisition).

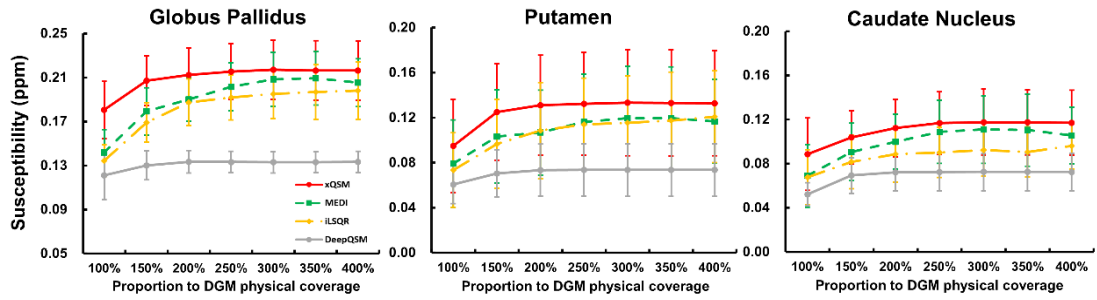

**Figure 7.** DGM (globus pallidus, caudate nucleus and putamen) measurements (mean and standard deviation) from three in vivo subjects. The effects of FOV coverage for dipole inversion are compared among different QSM methods.

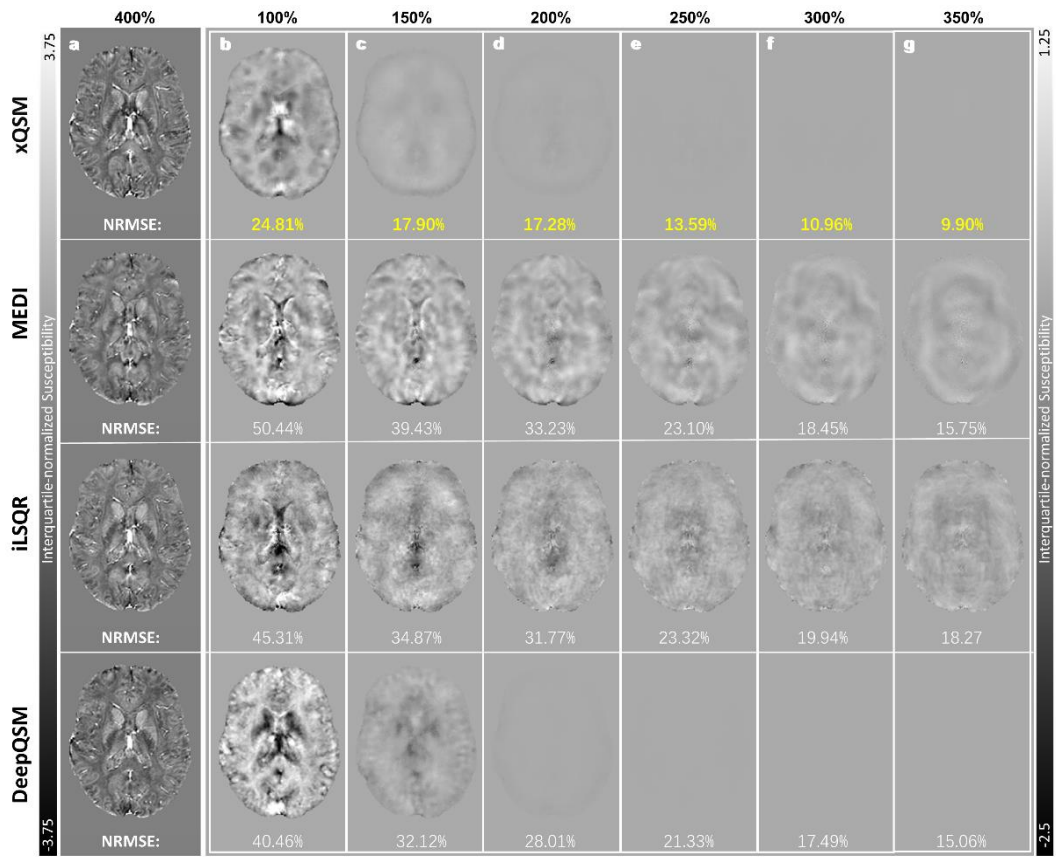

**Figure 8**. (a) Interquartile-normalized QSM from the full-brain FOV coverages (i.e., 400% of DGM coverage), and (b-g) difference maps of the smaller FOVs relative to the full-brain coverage of *In vivo* Subject #1. NRMSEs to the full-brain coverage are reported with the smallest values highlighted in bold yellow.

**Supplementary**

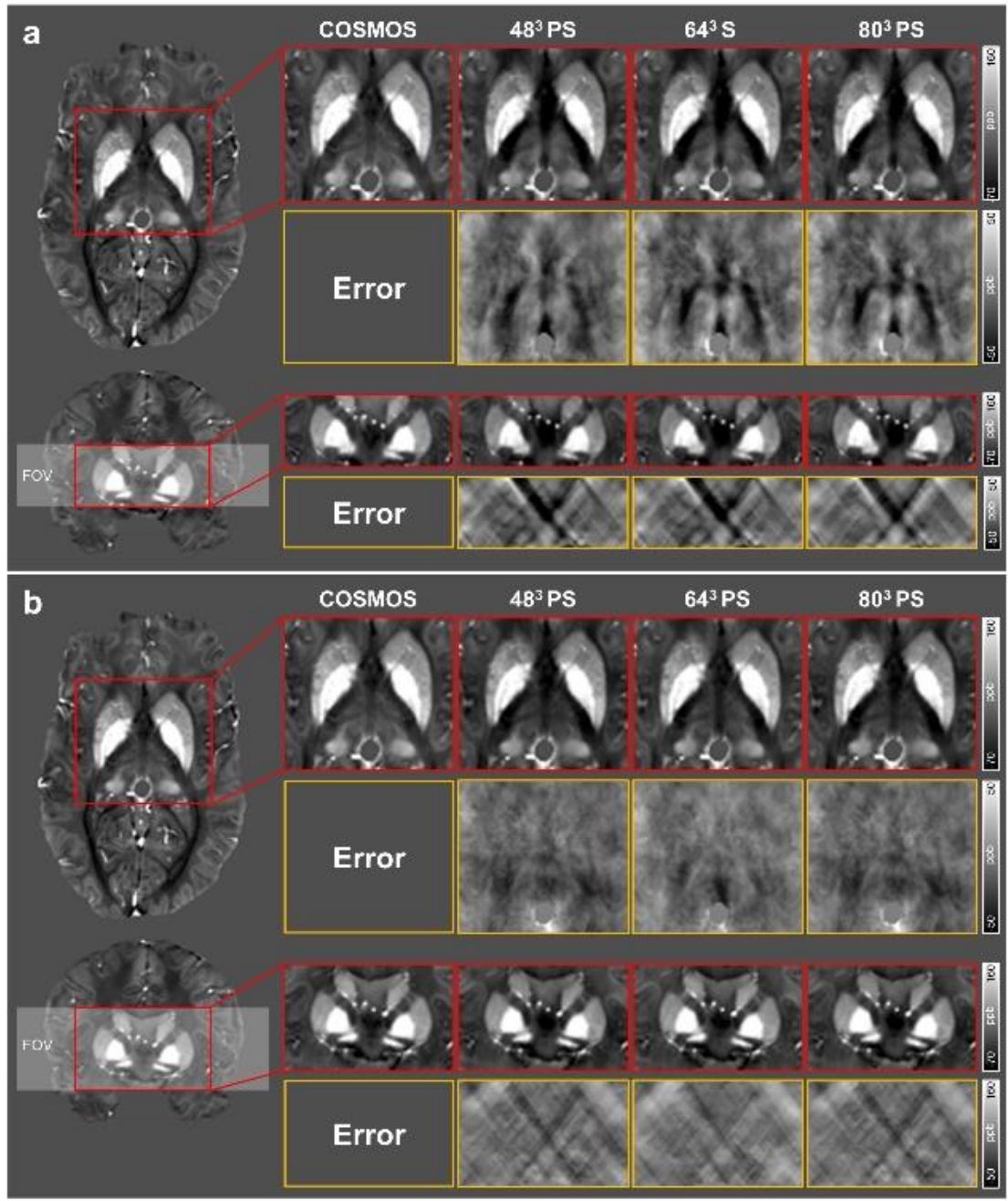

**Supplementary Figure 1:** Comparison of QSM reconstructions on one simulated subject from three different xQSM networks trained with patch sizes (PSs) of $48^3$, $64^3$, and $80^3$, under (a) 100% (i.e., 32 mm) and (b) 150% (i.e., 48 mm) DGM-FOV coverage. Error maps relative to the ground truth (COSMOS) are shown below the reconstructions. The grey shaded areas in (a) and (b) on the coronal images indicate different FOV coverages.

**Supplementary Table 1:** Error metrics of reconstructions from xQSM trained with different Patch Sizes.

| Patch Size: | | $48^3$ | $64^3$ | $80^3$ |
|---|---|---|---|---|
| 32 Slices (100% DGM) | SSIM | 0.77 | 0.79 | 0.77 |
| | PSNR | 37.68 | 38.65 | 37.61 |
| 48 Slices (150% DGM) | SSIM | 0.79 | 0.81 | 0.79 |
| | PSNR | 38.74 | 39.55 | 38.74 |
| 128 Slices (400% DGM) | SSIM | 0.87 | 0.89 | 0.85 |
| | PSNR | 41.59 | 42.82 | 40.42 |

The best results are highlighted in red.